\documentclass[12pt,showpacs,preprintnumbers,amsmath,aps,amssymb,epsfig]{revtex4}
\usepackage{graphicx}
\usepackage{dcolumn}
\usepackage{bm}
\usepackage{color}
\begin{document}

\title{Discrete canonical analysis of three dimensional gravity with cosmological constant}
\author{J. Berra-Montiel}
\email{jberra@fc.uaslp.mx} \affiliation{Facultad de Ciencias,
Universidad Aut\'onoma de San Luis Potos\'i, Av. Salvador Nava
S/N, Zona Universitaria CP 78290 San Luis Potos\'i, SLP, M\'exico}
\author{J.E. Rosales-Quintero}
\email{erosales@fisica.ugto.mx} \affiliation{Departamento de
F\'isica, DCI, Campus Le\'on, Universidad   de Guanajuato, A.P.
E-143, C.P. 37150, Le\'on, Guanajuato, M\'exico.}

\begin{abstract}
We discuss the interplay between standard canonical analysis and
canonical discretization in three-dimensional gravity with
cosmological constant. By using the Hamiltonian analysis, we find
that the continuum local symmetries of the theory are given by the
on-shell space-time diffeomorphisms, which at the action level,
corresponds to the Kalb-Ramond transformations. At the time of
discretization, although this symmetry is explicitly broken, we
prove that the theory still preserves certain gauge freedom
generated by a constant curvature relation in terms of holonomies
and the Gauss's law in the lattice approach.

\end{abstract}
\date{\today}
\pacs{00} \keywords{}

\maketitle

\section{ INTRODUCTION}

The construction of a consistent theory of quantum gravity is a
problem in theoretical physics that has so far defied all attempts
at resolution. The problem of finding a consistent quantum theory
of gravity is in part due to that general relativity is a very
complicated mathematical non-linear field theory and furthermore,
it's a geometric theory of spacetime, and quantizing gravity means
quantizing spacetime itself, and so far we do not know what this
means \cite{Kuchar}.

In order to overcome such difficulties, it is natural to look for
simpler models that share, mathematically   and physically, the
important conceptual features of general relativity, while
avoiding some of the computational difficulties. For quantum
gravity, a model as such, is general relativity in three
space-time dimensions. As a generally covariant field theory of
space-time geometry, $(2+1)$-dimensional gravity possesses the
same conceptual foundations as a realistic $(3+1)$-dimensional
general relativity, since many of the fundamental issues of
quantum gravity carry over to the lower dimensional setting
\cite{Witten},\cite{Carlip}.

For understanding fundamental concepts: the nature of time, the
construction of states and observables, the role of topology, and
the relationships among different quantization approaches, the
model has proven highly instructive \cite{Carlip}. Classical
solutions to the vacuum field equations turn out to be all locally
diffeomorphic to spacetimes of constant curvature, that is,
Minkowski, de Sitter, or anti-de Sitter spaces. This means that
any solution of the field equations with a cosmological constant
has constant curvature. Physically, a $(2+1)$-dimensional
spacetime has no local degrees of freedom, i.e., there are no
gravitational waves in the classical theory, and no propagating
gravitons in the quantum theory.

One of the central problems generated by the application of the
rules of quantum mechanics to a covariant field theory is the
problem of the dynamics. When formulated canonically, general
relativity has a vanishing Hamiltonian, which has to be
implemented as a constraint. In the loop quantum gravity approach
\cite{Rovelli}, this has been achieved \cite{Thiemann}, but to
characterize the resulting theory in a way in which the dynamics
of general relativity is explicit remains a challenge. In order to
overpass these obstacles, there has long been the hope that
lattice methods could be used as a non-perturbative approach to
quantum gravity. This is in part based on the fact that lattice
methods had been quite successful in the treatment of quantum
chromodynamics  (QCD), not only in making the theory finite, but
also in making it computable. However, in lattice QCD there exist
regularization methods that are gauge invariant, whilst in the
gravitational context this is not the case \cite{Loll}. As soon as
one discretizes space-time one breaks the invariance under
diffeomorphisms, then lattice methods in the gravitational setting
face unique challenges.

To address these problems, it has been proposed a systematic
canonical treatment for discretizing covariant field theories,
both at a classical and quantum mechanical level. This methodology
is related to a discretization technique called variational
integrators \cite{Marsden},
\cite{Lew},\cite{Gambini1},\cite{Gambini2}. This method consists
in discretizing the action of the theory and working from it the
discrete equations of motion. Automatically, the latter are
generically guaranteed to be consistent. The resulting discrete
theories have unique features that distinguish them from the
continuum theories, although a satisfactory canonical formulation
can be found for them. The discrete theories do not have
constraints associated with the space-time diffeomorphisms and as
a consequence the quantities that in the continuum are the
associated Lagrange multipliers become regular variables of the
discrete theories whose values are determined by the equations of
motion \cite{Gambini3}. Furthermore, the discretization breaks the
initial gauge freedom and solutions to the discrete theory, that
are different correspond, in the continuum limit, to the same
solution of the continuum theory. Hence the discrete theory has
more degrees of freedom. On the other hand, the lack of
constraints makes the discretized theories very promising at the
time of quantization, since most of the hard conceptual questions
of quantum gravity are related to the presence of constraints in
the theory \cite{Gambini4},\cite{Gambini5}.

Considering what is been stated, the purpose of this paper is to
present a consistent discretization of $(2+1)$-gravity with
cosmological constant, and discuss whether discretization leads to
a breaking of the local symmetries of the theory. The organization
of the paper is as follows: in section $2$, we study the
Kalb-Ramond transformations at the Lagrange level, in section $3$
we develop an extended canonical analysis of the continuum theory
by turning the quantities that play the role of Lagrange
multipliers into dynamical variables. This extended version,
although completely equivalent to the usual one, will allow us to
make a cleaner conection with the discrete theory. In section $4$
we briefly review the consistent discretization technique, and in
section $5$, we finally discuss $(2+1)$-gravity in a lattice. We
end with a discussion and present some conclusions.


\section{ Lagrangian formulation}
Let us take the Lie group $SU(2)$ as our gauge group. This group
is semisimple so there is  an invariant non degenerate bilinear
form, the so called, Cartan-Killing form. We take as our spacetime
a 3-dimensional oriented smooth manifold $\mathcal{M}$ with
$\mathcal{M}=\mathbb{R}\times \Sigma$, right by $\Sigma$ (which we
take to be compact and without boundary) corresponding to Cauchy's
surfaces and $\mathbb{R}$ representing an evolution parameter
(global hyperbolicity is imposed to exclude spacetimes with bad
causality properties). Now choose a principal $SU(2)$-bundle $P$
over $M$. Due to the fact that $SU(2)$ is simply connected,  this
$SU(2)$-bundle admits a trivialization \cite{Baez-bf}
\cite{Cattaneo BF 3-4 dim}. We can define the dynamical fields of
our theory as follows
\begin{itemize}
    \item A connection $A$ which is a $su(2)$-valued 1-form on
    $\mathcal{M}$, $A=A^{\ I}_{\mu}t_{I}dx^{\mu}$.
    \item A $su(2)$-valued 1-form field $e$ on $\mathcal{M}$, the dreibein, $e=e^{\ I}_{\mu}t_{I}dx^{\mu}$,
\end{itemize}
where we have defined $t_{I}$, with $I=1,2,3$, as the generators
of the $su(2)$ Lie algebra in the adjoint representation. These
generators satisfy
\begin{equation}
[t_{I},t_{J}]=2\epsilon_{IJ}^{\ \ \ K}t_{K}
\end{equation}
where $\epsilon_{IJK}$ is the totally antisymmetric tensor,
Levi-Civita tensor. The Cartan-Killing form is given by
\begin{equation}
\kappa_{IJ}=Tr(t_{I}t_{J})=\delta_{IJ}.
\end{equation}
Then, the action principle is given by
\begin{equation}\label{action}
S\left[ e,A\right]=\int_{\mathcal{M}}\ Tr \bigg[ e\wedge F
+\frac{\Lambda}{3}\ e\wedge  e\wedge e
\bigg]=\int_{\mathcal{M}}e_{I}\wedge F^{I}+\frac{\Lambda}{3}\
\epsilon_{IJK}e^{I}\wedge e^{J}\wedge e^{K},
\end{equation}
where $\Lambda$ is the cosmological constant and the  strength
tensor is a $su(2)$-valued 2-form which, as usual, it is taken as
$F=dA+A \wedge A$.  \\
The equations of motion from the action read
\begin{eqnarray}
 \label{1acartan }         \delta_{A}S=0 & \Rightarrow & De^{I}=0, \\
 \label{Einstein equations} \delta_{e}S=0 & \Rightarrow & F^{I}+\Lambda \ \epsilon^{IJK} e_{J}\wedge e_{K}=0,
\end{eqnarray}
where the covariant derivative is defined as,
$D\xi^{I}=d\xi^{I}+[A,\xi]^{I}$, for an $su(2)$-valued k-form
$\xi^{I}$. The spacetime is locally either flat ($\Lambda = 0$),
de Sitter ($\Lambda > 0$), or anti-de Sitter ($\Lambda < 0$).\\
 The equation (\ref{1acartan }) is the  zero-torsion condition, which let us
write the connection as function of the dreibein, $A=A(e)$. Then
when we use the last result back to the equation of motion
(\ref{Einstein equations}), we obtain
Riemannnian general relativity with cosmological constant \cite{Baez-bf}, \cite{4-dim BF Functor}.\\
The action, (\ref{action}), is invariant under $SU(2)$
transformations,
\begin{equation}
\delta_{\alpha}e=[e,\alpha] \qquad \delta_{\alpha}A=D\alpha.
\end{equation}
\noindent Inspired in the Kalb-Ramond transformation and the
transformation of the fields in Horowitz's theory for the BF model
\cite{Horowitz},\cite{Broda},\cite{Cattaneo BF 3-4 dim}, we find
that the action is invariant under field-dependent transformation
parameters, defined as
\begin{eqnarray}
  \delta_{C} e &=& DC, \\
  \delta_{C} A &=& \Lambda[C,e],
\end{eqnarray}
where $C$ are 0-form $su(2)$-valued transformation parameters.
These parameters generate  on-shell gauge transformations plus
diffeomorphisms of the basic fields. In this manner, if we define
the field-dependent parameters as $C^{I}=i_{v}e^{I}=v^{\mu}e^{\
I}_{\mu}$, the Kalb-Ramond transformations read
\begin{eqnarray}
    \delta_{v} e^{\ I}_{\nu} &=& D_{\nu}v^{\mu}\cdot e^{\ I}_{\mu}+v^{\mu}\cdot D_{\nu}e^{\
    I}_{\mu},\\
    \delta_{v} A^{\ I}_{\nu} &=& 2 v^{\mu}( \Lambda\epsilon^{IJK}e_{\mu J}e_{\nu K}).
\end{eqnarray}
Considering the equations of motion (\ref{1acartan }),
(\ref{Einstein equations}) and the Bianchi's identity $DF=0$, we
obtain
\begin{eqnarray}
  \delta_{v} e &=& \mathcal{L}_{v}e-\delta_{\alpha}e \\
  \delta_{v} A &=& \mathcal{L}_{v}A-\delta_{\alpha}A
\end{eqnarray}
where we have taken the $SU(2)$  field-dependent transformation
parameters as, $\alpha^{I}=v^{\rho}A_{\rho}^{\ I}$, and
$\mathcal{L}_{v}$ is the usual Lie derivative along the vector
field $v$. We can observe that these results coincide with the
fact that the previous action, (\ref{action}), is invariant under
diffeomorphisms and
internal gauge transformations by construction.\\
We now turn our attention to develop a canonical analysis of the
theory in order to show that the symmetries, we have found in the
Lagrangian formalism, are preserved at the Hamiltonian level.

\section{Canonical Analysis for the three dimensional gravity model}

In this section, we carry out an extended canonical analysis of
$(2+1)$-gravity; as we mentioned above, this analysis consists in
turning the quantities that play the role of Lagrange multipliers
into dynamical variables \cite{tesis},
\cite{Berra1},\cite{Berra2},\cite{Escalante}. This procedure, will
allow us to make a cleaner contact with the discrete version, as
in order to obtain a consistent discretization, some of the
Lagrange multipliers get determined by the scheme, and the
evolution is implemented by a canonical transformation, this means
that the set of discrete equations that were formely incompatible
can be solved simultaneously.

  We start from the action
(\ref{action}), taking both $e$ and $A$ as dynamical variables. By
performing the $2+1$ decomposition, we can write the action as
\begin{equation}
S\left[e,A
\right]=\int_{\mathit{R}\times\Sigma}\eta^{ij}\left[e_{0I}F^{\ \
I}_{ij}-2e_{iI}F_{0j}^{\ \ I}+\Lambda\epsilon_{IJK}e^{\ I}_{0}e^{\
J}_{i}e^{\ K}_{j} \right],
\end{equation}
\noindent where $\eta^{ij}=\epsilon^{0ij}$. From this action, we
identify the Lagrangian density
\begin{equation}
\mathcal{L}=\eta^{ij}\left[e_{0I}F^{\ \ I}_{ij}-2e_{iI}F_{0j}^{\ \
I}+\Lambda\epsilon_{IJK}e^{\ I}_{0}e^{\ J}_{i}e^{\ K}_{j} \right].
\end{equation}

\noindent By determining the set of dynamical variables, we need
the definition of the momenta $\left(p^{\alpha}_{\ A},
\pi^{\alpha}_{\ A} \right) $,
\begin{equation}
p^{\alpha}_{\ A}=\frac{\delta\mathcal{L}}{\delta \dot{e}^{\
A}_{\alpha}},
 \quad\quad \pi^{\alpha}_{\ A}=\frac{\delta\mathcal{L}}{\delta\dot{A}^{\ A}_{\alpha}},
\end{equation}

\noindent canonically conjugate to $\left(e^{\ A}_{\alpha},A^{\
A}_{\alpha} \right) $. The matrix elements of the Hessian,
\begin{equation}
\frac{\partial^2{\mathcal{L}} }{\partial (\partial_\mu A_{\alpha
}^{\ A} )\partial(\partial_\mu A_{\beta }^{\ B} ) }, \quad
 \frac{\partial^2{\mathcal{L}} }{\partial (\partial_\mu e^{\ A}_{\alpha} )
 \partial(\partial_\mu A_{\beta }^{\ B} ) }, \quad \frac{\partial^2{\mathcal{L}} }
 {\partial (\partial_\mu e^{\ A}_{\alpha} ) \partial(\partial_\mu e^{\ B}_{\beta } ) },
\end{equation}

\noindent vanish, which means that the rank of the Hessian is
equal to zero, so that,  18 primary constraints are expected. From
the definition of the momenta, it is possible to identify the
following 18 primary constraints:

\begin{eqnarray}
\phi^{0}_{\ A}&:& p^{0}_{\ A} \approx 0, \nonumber \\
\phi^{a}_{\ A}&:& p^{a}_{\ A} \approx 0, \nonumber \\
\psi^{0}_{\ A}&:& \pi^{0}_{\ A} \approx 0, \nonumber \\
\psi^{a}_{\ A}&:& \pi^{a}_{\ A}-2\eta^{ab}e_{b A} \approx 0.
\end{eqnarray}

\noindent By neglecting the terms on the frontier, the canonical
Hamiltonian for the three dimensional gravity model is expressed
as
\begin{equation}
H_{\rm c}=-\int_{\Sigma}d^{3}x\ \eta^{ij}\left[- e_{0I}F^{\ \
I}_{ij}-\Lambda\epsilon_{IJK}e^{\ I}_{0}e^{\ J}_{i}e^{\
K}_{j}\right]-A_{0}^{\ I}D_{i}\pi^{i}_{\ I}.
\end{equation}

\noindent By adding the primary constraints to the canonical
Hamiltonian, we obtain the primary Hamiltonian
\begin{equation}
H_{\rm P}=H_{\rm c}+\int_{\Sigma}d^{3}x \left[ \lambda_{0}^{\
I}\phi^{0}_{\ I} +\lambda^{\ I}_{i}\phi^{i}_{\ I}+\rho^{\
I}_{0}\psi^{0}_{\ I}+\rho^{\ I}_{i}\psi^{i}_{\ I}\right],
\end{equation}

\noindent where $\lambda^{\ I}_{0}$, $\lambda^{\ I}_{i}$, $\rho^{\
I}_{0}$ and $\rho^{\ I}_{i}$ are Lagrange multipliers enforcing
the constraints. The non-vanishing fundamental Poisson brackets
for the theory under study are given by
\begin{eqnarray}
\{e^{\ A}_{\alpha}(x^{0},x),  p^{ \mu}_{\ I}(y^{0},y)  \} & =& \delta^{\mu}_\alpha \delta^{A}_{ I}  \delta^2(x,y), \nonumber \\
\{ A^{\ A}_{\alpha}(x^{0},x), \pi^{\mu}_{\ I}(y^{0},y) \} &=&
\delta^\mu_\alpha \delta^{A}_{I} \delta^2(x,y).
\end{eqnarray}

\noindent Now, we need to identify if the theory has secondary
constraints. For this aim, we compute the $18\times 18$ matrix
whose entries are the Poisson brackets among the primary
constraints

\begin{eqnarray}\label{primary}
\{ \phi^{0}_{\ A}(x),\phi^{0}_{\ I}(y) \}&=&0,   \qquad  \{ \phi^{a}_{\ A}(x),\phi^{0}_{\ I}(y) \} = 0  \nonumber \\
\{ \phi^{0}_{\ A}(x),\phi^{i}_{\ I}(y) \} &=& 0, \qquad  \{ \phi^{a}_{\ A}(x),\phi^{i}_{\ I}(y) \} = 0, \nonumber \\
\{ \phi^{0}_{\ A}(x),\psi^{0}_{\ I}(y) \} &=& 0, \qquad  \{ \phi^{a}_{\ A}(x),\psi^{0}_{\ I}(y) \} = 0, \nonumber \\
\{ \phi^{0}_{\ A}(x),\psi^{i}_{\ I}(y) \} &=& 0, \qquad  \{ \phi^{a}_{\ A}(x),\psi^{i}_{\ I}(y) \} =- 2\eta^{ai}\delta_{AI}\delta(x,y), \nonumber \\
\{ \psi^{0}_{\ A}(x),\phi^{0}_{\ I}(y) \}&=&0,   \qquad  \{ \psi^{a}_{\ A}(x),\phi^{0}_{\ I}(y) \} = 0  \nonumber \\
\{ \psi^{0}_{\ A}(x),\phi^{i}_{\ I}(y) \} &=& 0, \qquad  \{ \psi^{a}_{\ A}(x),\phi^{i}_{\ I}(y) \} = -2\eta^{ai}\delta_{AI}\delta(x,y), \nonumber \\
\{ \psi^{0}_{\ A}(x),\psi^{0}_{\ I}(y) \} &=& 0, \qquad  \{ \psi^{a}_{\ A}(x),\psi^{0}_{\ I}(y) \} = 0, \nonumber \\
\{ \psi^{0}_{\ A}(x),\psi^{i}_{\ I}(y) \} &=& 0, \qquad  \{ \psi^{a}_{\ A}(x),\psi^{i}_{\ I}(y) \} =0. \nonumber \\
\end{eqnarray}

\noindent This matrix has rank=12 and 6 linearly independent
null-vectors, which implies that there are 6 secondary constraints
to be determined by consistency conditions. By requiring
consistency of the temporal evolution of the constraints and the 6
null vectors, 6 secondary constraints arise,
\begin{eqnarray}
\dot{\phi}^{0}_{\ A}&=&\{\phi^{0}_{A}, H_{P} \}\approx 0 \quad \Rightarrow \quad D^{A}:= \eta^{ab}F_{ab}^{\ \ A}+\Lambda\eta^{bc}\epsilon^{A}_{\;\;BC}e^{\ B}_{b}e^{\ C}_{c} \approx 0,  \nonumber \\
\dot{\psi}^{0}_{\ A}&=&\{\psi^{0}_{A}, H_P \}\approx 0 \quad
\Rightarrow \quad G_{A}:= D_{a}\pi^{a}_{\ A} \approx 0,
\end{eqnarray}

\noindent and the following Lagrange multipliers are fixed,
\begin{eqnarray}\label{multipliers}
\dot{\phi}^{a}_{\ A}&=&\{\phi^{a}_{\ A}, H_P \} \approx 0 \quad \Rightarrow \quad \rho_{aA}=-\Lambda\epsilon_{ABC}e^{\ B}_{0}e^{\ C}_{a},  \nonumber \\
\dot{\psi}^{a}_{\ A}&=&\{\psi^{a}_{\ A}, H_P \}\approx 0 \quad
\Rightarrow \quad \lambda^{\ A}_{a}=-D_{a}e^{\
A}_{0}+2\epsilon^{A}_{\;\;BC}A^{\ B}_{a}e^{\ C}_{0}-\frac{1}{2}\
\epsilon^{A}_{\;\;BC}A^{\ B}_{0}\pi_{a}^{\ C} .
\end{eqnarray}
\noindent This theory does not have terciary constraints. By
following the method, we determine which of the constraints
(primary and secondary) are first class and which are second
class. To accomplish such a task we calculate the Poisson brackets
between the primary and secondary constraints. To complete the
constraint matrix, we add to the algebra shown in Eq.
(\ref{primary})  the following expressions
\begin{eqnarray}\label{secondary}
\{ \phi^{0}_{\ A}(x),G_{I}(y) \}&=&0,   \qquad  \{ \phi^{a}_{\ A}(x),G_{I}(y) \} = 0  \nonumber \\
\{ \phi^{0}_{\ A}(x),D^{I}(y) \} &=& 0, \qquad  \{ \phi^{a}_{\ A}(x),D^{I}(y) \} = 2\Lambda\eta^{ab}\epsilon^{I}_{\ AB}e_{b}^{\ B}\delta(x,y), \nonumber \\
\{ \psi^{0}_{\ A}(x),G_{I}(y) \}&=&0,   \qquad  \{ \psi^{a}_{\ A}(x),G_{I}(y) \} = \epsilon_{AI}^{\;\;\;\;K}\pi^{a}_{\ K}(y)\delta(x,y)  \nonumber \\
\{ \psi^{0}_{\ A}(x),D^{I}(y) \} &=& 0, \qquad  \{ \psi^{a}_{\ A}(x),D^{I}(y) \} =2\eta^{ai}\left[\delta^{I}_{A}\partial_{i}(y)+\epsilon_{A\;K}^{\;\;\;I}A^{\ K}_{i}(y)\right]\delta(x,y), \nonumber \\
\{ D^{A}(x),D^{I}(y) \}&=&0,   \qquad   \{ G_{A}(x),D^{I}(y) \} = \epsilon_{A}^{\;\ IC}D_{C}\delta(x,y)=0  \nonumber \\
\{ G_{A}(x),G_{I}(y)
\}&=&\epsilon_{AI}^{\;\;\;\;C}G_{C}\delta(x,y)=0.
\end{eqnarray}
\noindent The matrix formed by the Poisson brackets among all the
constraints exhibited in Eqs. (\ref{primary}) and
(\ref{secondary}) has rank=18 and 6 null-vectors. The contraction
of the null vectors with the matrix formed by the constraints
results in 12 first class constraints:

\begin{eqnarray}\label{firstclass}
\phi^{0}_{\ A}&:& p^{0}_{\ A}, \nonumber \\
\psi^{0}_{\ A}&:& \pi^{0}_{\ A}, \nonumber \\
 G_{A}&:& D_{a}\pi^{a}_{\ A}+\epsilon_{AB}^{\;\;\;\; \;C}e^{\ B}_{a}p^{a}_{\ C},\nonumber \\
 D^{A}&:&\eta^{ab}F_{ab}^{\ \  A}-\Lambda\eta^{ab}\epsilon^{A}_{\;\;BC}e^{\ B}_{a}e^{\ C}_{b}+\Lambda\epsilon^{A\;C}_{\;B}e^{\ B}_{a}\pi^{a}_{\ C}+D_{a}p^{aA}.
\end{eqnarray}

On the other hand, we find 12 second class constraints:
\begin{eqnarray}\label{secondclass}
\phi^{a}_{\ A}&:& p^{a}_{\ A}, \nonumber \\
\psi^{a}_{\ A}&:& \pi^{a}_{\ A}-\eta^{ab}e_{b A}.
\end{eqnarray}
\noindent After this analysis, we conclude the model has 18
canonical variables, 12 independent first class constraints and 12
independent second class constraints, which leads to determine, by
performing a counting of the degrees of freedom \cite{Henneaux},
that this model has none degrees of freedom per space-time point.
Of course, by considering the second class constraints Eq.
(\ref{secondclass}) as strong equations, the above relations are
reduced to the usual constraints \cite{Witten}, so that this
analysis extends and completes the results in the literature.

\noindent By calculating the algebra among the constraints, we
find that

\begin{eqnarray}\label{algebra}
\{ \phi^{0}_{\ A}(x),\phi^{0}_{\ I}(y) \}&=&0,   \qquad   \{ \phi^{a}_{\ A}(x),\phi^{i}_{\ I}(y) \} = 0, \nonumber \\
\{ \phi^{0}_{\ A}(x),\phi^{i}_{\ I}(y) \} &=& 0, \qquad   \{ \phi^{a}_{\ A}(x),\psi^{0}_{\ I}(y) \} = 0, \nonumber \\
\{ \phi^{0}_{\ A}(x),\psi^{0}_{\ I}(y) \} &=& 0, \qquad   \{ \phi^{a}_{\ A}(x),\psi^{i}_{\ I}(y) \} =-2\eta^{ai}\delta_{AI}\delta(x,y), \nonumber \\
\{ \phi^{0}_{\ A}(x),\psi^{i}_{\ I}(y) \} &=& 0, \qquad   \{ \phi^{a}_{\ A}(x),G_{I}(y) \} = \epsilon_{AI}^{\;\;\; \; K}\phi^{a}_{\ K}\delta(x,y), \nonumber \\
\{ \phi^{0}_{\ A}(x),G_{I}(y) \} &=& 0,          \qquad   \{ \phi^{a}_{\ A}(x),D^{I}(y) \} = 0, \nonumber \\
\{ \phi^{0}_{\ A}(x),D^{I}(y) \} &=& 0,          \qquad   \{ \psi^{a}_{\ A}(x),\psi^{i}_{\ I}(y) \}= 0,  \nonumber \\
\{ \psi^{0}_{\ A}(x),\psi^{0}_{\ I}(y) \}&=&0,   \qquad   \{ \psi^{a}_{\ A}(x),G_{I}(y) \}=\epsilon_{AI}^{\;\;\;\;\; K}\psi^{a}_{\ K}\delta(x,y),\nonumber \\
\{ \psi^{0}_{\ A}(x),\psi^{i}_{\ I}(y) \}&=&0,   \qquad   \{ \psi^{a}_{\ A}(x),D^{I}(y) \} = \epsilon_{A\; K}^{\;\;\;I}\phi^{a}_{\ K}\delta(x,y),  \nonumber \\
\{ \psi^{0}_{\ A}(x),G_{I}(y) \}&=&0,            \qquad   \{ G_{A}(x),G_{I}(y) \}=\epsilon_{AI}^{\;\;\;\;C}G_{C}\delta(x,y), \nonumber \\
\{ \psi^{0}_{\ A}(x),D^{I}(y) \}&=&0,            \qquad   \{ G_{A}(x),D^{I}(y) \}=\epsilon_{A\;C}^{\;\;\;I}D^{C}\delta(x,y), \nonumber\\
\{ D^{A}(x),D^{I}(y) \}&=& 0,
\end{eqnarray}

\noindent from where we can appreciate that the constraints form a
set of first and second class constraints, as expected. The
determination of the nature of the constraints allow us to find
the extended action. By employing the first class constraints, Eq.
(\ref{firstclass}), the second class constraints, Eq.
(\ref{secondclass}), and the Lagrange multipliers, Eq.
(\ref{multipliers}), we find that the extended action takes the
form
\begin{equation}
S_{E}[e,p,A,\pi,\lambda,\rho,\gamma,\xi] = \int \left[
\dot{e}_{\alpha}^{\ A}p^{\alpha}_{\ A}+\dot{A}_{\alpha}^{\
A}\pi^{\alpha}_{\ A}-H -\lambda_{\alpha}^{\ A}\phi^{\alpha}_{\
A}-\rho^{\ A}_{\alpha}\psi^{\alpha}_{\
A}-\gamma^{A}G_{A}-\xi_{A}D^{A}\right]d^{3}x,
\end{equation}
\noindent where $H$ is a linear combination of first class
constraints, and is given by

\begin{equation}\label{Hamiltonian}
H=-A_{0}^{\ A}G_{A}-e_{0A}D^{A}
\end{equation}

\noindent and $\lambda_{\alpha}^{\ A}$ , $\rho_{\alpha}^{\ A}$ ,
$\gamma^{A}$ , $\xi_{A}$ are the Lagrange multipliers enforcing
the first and second class constraints. We observe, by considering
the second class constraints as strong equations, that the
Hamiltonian shown in Eq. (\ref{Hamiltonian}) is reduced to the
usual expression found in the literature \cite{Witten},
\cite{Carlip}, which is defined on a reduced phase space context.
From the extended action, we identify the extended Hamiltonian,
$H_{\rm E}$, which is given by
\begin{equation}\label{extendedH}
H_{\rm E}=H-\lambda_{0}^{\ A}\phi^{0}_{\
A}-\rho^{A}_{0}\psi^{0}_{A}-\gamma^{A}G_{A}-\xi_{A}D^{A}.
\end{equation}
\noindent By utilizing our expressions for the complete set of
constraints, it is possible to obtain the gauge transformations
acting on the full phase space. For this important step, we shall
use the Castellani's formalism \cite{Castellani}, which allows us
to define the following gauge generator in terms of the first
class constraints:
\begin{equation}\label{generator}
G=\int_{\Sigma}\left[ D_{0}\varepsilon_{0}^{\
A}\phi^{0}_{A}+D_{0}\zeta_{0}^{\ A}\psi^{0}_{\
I}+\varepsilon^{A}G_{A}+\zeta_{A}D^{A}\right]d^{3}x,
\end{equation}
\noindent where $\varepsilon_{0}^{\ A}$, $ \varepsilon^{\ A}$,
$\zeta_{0}^{\ A}$ and $\zeta_{A}$ are arbitrary continuum real
parameters. Thus, we find that the gauge transformations in the
phase space are
\begin{eqnarray}
\delta_{0}e_{0}^{\ A}&=& D_{0}\varepsilon_{0}^{\ A}, \nonumber \\
\delta_{0}e_{a}^{\ A}&=&\epsilon^{A}_{\;\;IJ}\varepsilon^{I}e^{\ J}_{a}-D_{a}\zeta^{A}, \nonumber \\
\delta_{0}A_{0}^{\ A}&=& D_{0}\zeta_{0}^{\ A}, \nonumber \\
\delta_{0}A_{a}^{\ A}&=& -D_{a}\varepsilon^{A}+\Lambda\epsilon^{AI}_{\;\;\;J}e^{\ J}_{a}, \nonumber \\
\end{eqnarray}

\noindent In order to recover the on-shell diffeomorphisms
symmetry, one can redefine the gauge parameters as
$-\varepsilon_{0}^{\ I}=\varepsilon^{I}=-v^{\alpha}A_{\alpha}^{\
I}$, and $-\zeta^{\ I}_{0}=\zeta^{I}=-v^{\alpha}e_{\alpha}^{\ I}$.
By such an election, the gauge tranformations take the form
\begin{eqnarray}\label{gauge}
e_{\alpha}^{'A} &\longrightarrow & e_{\alpha}^{A}+\mathcal{L}_{v}e^{A}_{\alpha}+\left(v\times De\right)^{\ A}_{\alpha}, \nonumber \\
A_{\alpha}^{'A} &\longrightarrow &
A_{\alpha}^{A}+\mathcal{L}_{v}A^{I}_{\alpha}+\left[v\cdot\left(F+\Lambda
e\wedge e\right)\right]^{\ A}_{\alpha}.
\end{eqnarray}
Where "$\times$", and "$\cdot$" denote the usual cross and dot
product in the three dimensional space-time. By examining the
constraints in the complete phase space, we have obtained, in an
explicit form, the generators of the gauge transformations for all
fields within the action, even if they behave like Lagrange
multipliers in accordance with the on-shell Kalb-Ramond
transformations found in the last section.

\section{Consistent discretization of constrained theories}

We illustrate the technique with a mechanical system for
simplicity, the case of field theories is straightforward, since
upon discretization the latter become mechanical systems
\cite{Marsden},\cite{Gambini1},\cite{Gambini2}. We assume we start
from an action in the continuum, written in first-order form,
\begin{equation}
S=\int L(q,p)dt,
\end{equation}
with
\begin{equation}
L(q,p)=p\dot{q}-H(q,p)-\lambda_{B}\phi^{B}(q,p),
\end{equation}
where $\lambda_{B}$ are the Lagrange multipliers that enforce the
constraints $\phi^{B}(q,p)=0$. The discretization of the action
yields $S=\sum_{n=0}^{N}L(n,n+1)$, where
\begin{equation}
L(n,n+1)=p_{n}(q_{n+1}-q_{n})-\varepsilon
H(q_{n},p_{n})-\lambda_{nB}\phi^{B}(q_{n},p_{n}),
\end{equation}
where $\varepsilon=t_{n+1}-t_{n}$, and we have absorbed an
$\varepsilon$ in the definition of the Lagrange multipliers. In
the discrete setting, the Lagrangian can be seen has the generator
of a type 1 canonical transformations between the instant $n$ and
the instant $n+1$. In order to obtain a fully consistent theory,
 the equations of the discrete theory must be solved
simultaneously, we will view $q_{n}$, $p_{n}$, $\lambda_{nB}$ and
$q_{n+1}$, $p_{n+1}$, $\lambda_{n+1B}$ as configuration variables
and will assign to each of them a canonically conjugate momentum,
$P_{n}^{q}$, $P_{n}^{p}$, $P_{n}^{\lambda_{B}}$ and $P_{n+1}^{q}$,
$P_{n+1}^{p}$, $P_{n+1}^{\lambda_{B}}$. If one explicitly computes
the partial derivatives with the Lagrangian already given, one can
obtain a more familiar-looking set of equations,
\begin{eqnarray}
p_{n}-p_{n-1}&=&-\varepsilon\frac{\partial H(q_{n},p_{n})}{\partial q_{n}}-\lambda_{nB}\frac{\partial\phi^{B}(q_{n},p_{n})}{\partial q_{n}}, \nonumber\\
q_{n+1}-q_{n}&=&\varepsilon\frac{\partial H(q_{n},p_{n})}{\partial p_{n}}+\lambda_{nB}\frac{\partial\phi^{B}(q_{n},p_{n})}{\partial p_{n}},\\
\phi^{B}(q_{n},p_{n})&=&0 \nonumber.
\end{eqnarray}
These equations appear very similar to the ones one would obtain
by first working out the equations of motion in the continuum and
then discretizing them. A significant difference, however, is that
when one solves this set of equations some of the Lagrange
multipliers get determined, they are not free anymore as they are
in the continuum \cite{Sigma}. We therefore see that generically
when one discretizes constrained theories one gets a different
structure than in the continuum, in which some of the Lagrange
multipliers get undetermined. The equations that in the continuum
used to be constraints become upon discretization
pseudo-constraints in that they relate variables at different
instants of time and are solved by determining the Lagrange
multipliers.


\section{The consistent discretization approach}

In this section, we will apply the consistent discretization
technique to the three-dimensional gravity with cosmological
constant (\ref{action}), defined on a lattice. We discretize the
three dimensional gravity action as follows (see Figure 1),
\begin{eqnarray}
L(n,n+1)&&=\sum_{v}Tr [e^{0}_{n,v}h^{1}_{n,v}h^{2}_{n,v+e_{1}}(h^{1}_{n,v+e_{2}})^{\dagger}(h^{2}_{n,v})^{\dagger}+e^{2}_{n,v}V_{n,v}h^{1}_{n+1,v}(V_{n,v+e_{1}})^{\dagger}(h^{1}_{n,v})^{\dagger} \nonumber \\
&&-e^{1}_{n,v}V_{n,v}h^{2}_{n+1,v}(V_{n,v+e_{2}})^{\dagger}(h^{2}_{n,v})^{\dagger}+\rho_{n,v}(V_{n,v}(V_{n,v})^{\dagger}-1)+\lambda^{1}_{n,v}(h^{1}_{n,v}(h^{1}_{n,v})^{\dagger}-1) \nonumber \\
&&+\lambda^{2}_{n,v}(h^{2}_{n,v}(h^{2}_{n,v})^{\dagger}-1)+\Lambda\
e^{0}_{n,v}e^{1}_{n,v}e^{2}_{n,v}].
\end{eqnarray}

\noindent By $h^{a}_{n,v}$ we mean the holonomy along the spatial
direction of the elementary unit vector $e_{1}, e_{2}$, starting
from the lattice
 point labeled by the time step $n$ and the spatial point $v$ (which is labelled by a pair of indices as we are dealing with a three dimensional field theory). By the
 index $v+e_{a}$, we mean a lattice point coming from the spatial point $v$ in direction of $e_{a}$. The $e$ fields live in a
 one dimensional surface, dual to the plaquette on which we compute the holonomy representing the curvature field $F$, and
hence, it is an element of $SU(2)$, i.e,  an algebra valued one
form. The symbol $V_{n,v}$ represents the vertical holonomies and
are only labelled by
 the lattice point they start at. We will consider that the holonomies are matrices of the form $h=h^{A}t_{A}, V=V^{A}t_{A}$, where
 $t^{0}=I/\sqrt{2}$ and $t^{I}=-i\sigma^{I}/\sqrt{2}$, and $\sigma^{I}$ are the well known Pauli matrices (In this section, we consider the capital
 latin letters from the beginning of the alphabet as labels of the gauge group $A, B, \ldots =\{0, I\}$ ). Finally, $\rho$ and $\lambda$ are
 Lagrange multipliers that enforce, the above defined holonomies are indeed elements of $SU(2)$.

 The canonical variables of the theory are $h^{a,A}_{n,v}$, $V_{n,v}^{A}$, $e^{a, I}_{n,v}$, $e^{0, I}_{n,v}$,
$\rho_{n,v}$ and $\lambda^{a}_{n,v}$ where $a=1,2$.

\begin{figure}
  \centering
    \includegraphics[width=0.5\textwidth]{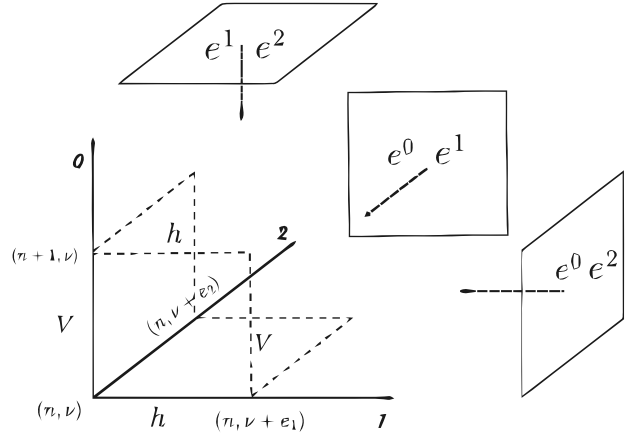}
  \caption{A elementary cell for a discrete description of 2+1 gravity.}
  \label{lattice}
\end{figure}

The momenta associated for $\lambda^{a}_{n,v}$, are given by
\begin{eqnarray}
\Pi^{\lambda^{a}}_{n+1,v}&\equiv&\frac{\partial L(n,n+1)}{\partial \lambda^{a}_{n+1,v}}=0, \nonumber \\
\Pi^{\lambda^{a}}_{n,v}&\equiv&-\frac{\partial L(n,n+1)}{\partial
\lambda^{a}_{n,v}}=-Tr\left[h^{a}_{n,v}(h^{a}_{n,v})^{\dagger}-1\right].
\end{eqnarray}

\noindent Consistency in time leads to the secondary class
constraint:
\begin{equation}
Tr \left[h^{a}_{n,v}(h^{a}_{n,v})^{\dagger}-1\right]=0,
\end{equation}

\noindent which, after time evolution, does not yield any new
constraint. The momenta associated for $\rho_{v,n}$,  are given by
\begin{eqnarray}
\Pi^{\rho}_{n+1,v}&\equiv&\frac{\partial L(n,n+1)}{\partial \rho_{n+1,v}}=0, \nonumber \\
\Pi^{\rho}_{n,v}&\equiv&-\frac{\partial L(n,n+1)}{\partial
\rho_{n,v}}=-Tr\left[V_{n,v}(V_{n,v})^{\dagger}-1\right].
\end{eqnarray}

\noindent Consistency in time leads to the secondary class
constraint
\begin{equation}
Tr\left[V_{n,v}(V_{n,v})^{\dagger}-1\right]=0,
\end{equation}

\noindent which, again is preserved in time without generating any
new constraint. Now, we consider the momenta associated to the
fields $e^{a}_{n,v}$, which read
\begin{eqnarray}\label{Fzero}
\Pi^{e^{a}}_{n+1,\nu}&\equiv& \frac{\partial L(n,n+1)}{\partial e^{a,I}_{n+1,v}}\ (T^{I})^\dagger =0, \nonumber \\
\Pi^{e^{a}}_{n,\nu}&\equiv&-\frac{\partial L(n,n+1)}{\partial
e^{a,I}_{n,v}}(T^{I})^\dagger=-Tr\left[h^{bc}_{n,v}T_{I}+\Lambda
e_{n,v}^{b}e_{n,v}^{c} T_{I}\right](T^{I})^\dagger,
\end{eqnarray}

\noindent where $\{a,b,c\}$ represents a cyclic permutation of $\{
0,1,2\}$, and $h^{ab}_{n,v}$ stands for an holonomy starting from
the point labelled by $n$, $v$, around the elementary plaquette in
the plane $a - b$. On the other hand, $e_{n,v}^{a}e_{n,v}^{b}$, is
the dual  plaquette related  to the $a - b$ holonomy, as shown in
figure (\ref{lattice}). Consistency in time of the constraint
(\ref{Fzero}) leads to
\begin{equation}
Tr\left[h^{ab}_{n,v}T_{I}+\Lambda\ e_{n,v}^{a}e_{n,v}^{b}
T_{I}\right]=0,
\end{equation}
\noindent the most general solution for this equation is given by
\begin{equation}  \label{eq: general solution traces}
h^{bc}_{n,v}=-\Lambda\ e^{b}_{n,v}e^{c}_{n,v}+\sigma^{bc}_{n,v}I
\end{equation}
where $(h^{bc}_{n,v})^{\dag}=h^{cb}_{n,v}$ and
$\sigma^{ab}_{n,v}=\pm 1$, this sign is arbitrary and can change
from  plaquette to plaquette. The appearance of $\sigma$ is due
that the solution is defined upon a traceless
term\cite{Gambini1}.\\
The expression for the momenta associated with the vertical
holonomies, $V_{n,v}^{A}$, are
\begin{eqnarray}\label{Vconstraints}
\Pi_{n+1,v}^{V}&=&\frac{\partial L(n,n+1)}{\partial V_{n+1,v}^{A}}(T^{A})^\dagger=0, \nonumber \\
V_{n,v}\Pi_{n,v}^{V}&=&h^{01}_{n,v}e^{2}_{n,v}-h^{02}_{n,v}e^{1}_{n,v}-(h^{1}_{n,v-e_{1}})^{\dagger}h^{10}_{n,v-e_{1}}e^{2}_{n,v-e_{1}}h^{1}_{n,v-e_{1}} \nonumber  \\
&+&(h^{2}_{n,v-e_{2}})^{\dagger}h^{20}_{n,v-e_{2}}e^{1}_{n,v-e_{2}}h^{2}_{n,v-e_{2}}+2\rho_{n,v}I.
\end{eqnarray}
\noindent Preserving in time the constraints (\ref{Vconstraints})
, together with (\ref{eq: general solution traces}), we obtain
\begin{equation}\label{defV}
2Tr(-\Lambda e^{0}_{n,v}e^{1}_{n,v}e^{2}_{n,v})+Tr(-\Lambda
e^{0}_{n,v-e_1}e^{1}_{n,v-e_1}e^{2}_{n,v-e_1})+Tr(-\Lambda
e^{0}_{n,v-e_2}e^{1}_{n,v-e_2}e^{2}_{n,v-e_2})=-4\rho_{n,v},
\end{equation}

\noindent where we can notice, that all terms are the same three
volume element, but evaluated at diferent points. Because the
cells are all equivalent and they do not depend on the label
attached to them, it is useful to define the trace of the volume
element as $\mathcal{V}=Tr(e^{0}_{n,v}e^{1}_{n,v}e^{2}_{n,v})$.
Taking this into account, consistency in time of the constraint
(\ref{Vconstraints}), allow us to fix the multiplier,
$\rho_{n,v}=\Lambda \mathcal{V}$. Then, the equation of motion
associated to the vertical holonomies, $V_{n,v}$, reads
\begin{equation}     \label{eq: of motion for Gauss's law}
h^{01}_{n,v}e^{2}_{n,v}-h^{02}_{n,v}e^{1}_{n,v}-(h^{1}_{n,v-e_1})^{\dag}h^{10}_{n,v-e_1}e^{2}_{n,v-e_1}h^{1}_{n,v-e_1}
+(h^{2}_{n,v-e_2})^{\dag}h^{20}_{n,v-e_2}e^{1}_{n,v-e_2}h^{2}_{n,v-e_2}=-2\Lambda
\mathcal{V} I,
\end{equation}

\noindent where $I$, denotes the identity matrix. In a similar
fashion, and in accordance with the consistent discretization
approach, the equation of motion associated to the spatial
holonomy, $h^{1}_{n+1,v}$, defined at the point, $(n+1,v)$, is
given by
\begin{equation}
h^{1}_{n+1,v}\Pi^{h^{1}}_{n+1,v}=(V_{n,v})^{\dag}h^{01}_{n,v}e^{2}_{n,v}V_{n,v},
\end{equation}
\noindent by using (\ref{eq: general solution traces}), we obtain
the primary constraint
\begin{equation}\label{electric1}
Tr(h^{1}_{n+1,v}\Pi^{h^{1}}_{n+1,v})=-\Lambda \mathcal{V}.
\end{equation}
\noindent Furthermore, the equation of motion associated with the
spatial holonomy, $h^{1}_{n,v}$, this time evaluated at the point
$(n,v)$, leads to
\begin{equation}
h^{1}_{n,v}\Pi^{h^{1}}_{n,v}=h^{10}_{n,v}e^{2}_{n,v}-h^{12}_{n,v}e^{0}_{n,v}+(h^{2}_{n,v-e_2})^{\dag}h^{21}_{n,v-e_2}e^{0}_{n,v-e_2}h^{2}_{n,v-e_2}
+2\lambda^{1}_{n,v}I,
\end{equation}
\noindent preserving in time this contraint, allows us to fix the
Lagrange multiplier
\begin{equation}
\lambda^{1}_{n,v}=-\Lambda \mathcal{V}.
\end{equation}
We proceed analogously for the variable $h^{2}_{n,v}$. The
equation of motion associated with the spatial holonomy
$h^{2}_{n+1,v}$, defined at the point $(n+1,v)$ is
\begin{equation}
h^{2}_{n+1,v}\Pi^{h^{2}}_{n+1,v}=-(V_{n,v})^{\dag}h^{02}_{n,v}e^{1}_{n,v}V_{n,v},
\end{equation}
\noindent from which we obtain the primary constraint
\begin{equation}\label{electric2}
Tr(h^{2}_{n+1,v}\Pi^{h^{2}}_{n+1,v})=-\Lambda \mathcal{V},
\end{equation}
\noindent and the equation of motion related to the spatial
holonomy, $h^{2}_{n,v}$, at the point $(n,v)$, gives us
\begin{equation}
h^{2}_{n,v}\Pi^{h^{2}}_{n,v}=h^{21}_{n,v}e^{0}_{n,v}-h^{20}_{n,v}e^{1}_{n,v}-(h^{1}_{n,v-e_1})^{\dag}h^{12}_{n,v-e_1}e^{0}_{n,v-e_1}h^{1}_{n,v-e_1}
+2\lambda^{2}_{n,v}I,
\end{equation}
\noindent which fix the corresponding Lagrange multiplier
\begin{equation}
\lambda^{2}_{n,v}=-\Lambda \mathcal{V}.
\end{equation}

We now consider the variable,
$E^{a}_{n,v}=h^{a}_{n,v}\Pi^{h^a}_{n,v}$ with $a\in \{1,2\}$, such
combination of spatial holonomies and their associated conjugate
momenta, play the role of an electric field, along the direction
$e_{a}$ at the point $(n,v)$, in the study of non-abelian theories
in the lattice approach. In our case, due to the constraints
(\ref{electric1}) and (\ref{electric2}), we observe that the
electric field $E^{a}_{n,v}$, is not an element of the algebra,
this occurs because, a closed loop  results modified by the
presence of the cosmological constant
\begin{equation}
Tr (E^{a}_{n,v})=-\Lambda \mathcal{V}.
\end{equation}
For our purposes, it is useful to define the electric field in the
opposite direction. $-e_{a}$, evaluated at the point $\{ n,v \}$,
as
\begin{equation}
E^{\bar{a}}_{n,v}=
h^{\bar{a}}_{n,v}\Pi^{h^{\bar{a}}}_{n,v}=(h^{a}_{n,v-e_a})^{\dag}(\Pi^{h^{a}}_{n,v-e_a})^{\dag}
=(h^{a}_{n,v-e_a})^{\dag}(E^{a}_{n,v-e_a})^{\dag}h^{a}_{n,v-e_a}.
\end{equation}
In order to calculate $(E^{a}_{n,v})^{\dagger}$, we make use of
the following straightforward relations, which  can be derived
directly from the constraint (\ref{eq: general solution traces}):
\begin{eqnarray}
        h^{cb}_{n,v}                  &=&      h^{bc}_{n,v}+\Lambda\ [e^{b}_{n,v},e^{c}_{n,v}]\\
        e^{a}_{n,v}(h^{bc}_{n,v})     &=&     (h^{bc}_{n,v})e^{a}_{n,v}+[e^{a}_{n,v},h^{bc}_{n,v}].
\end{eqnarray}
\noindent Therefore, the hermitian conjugate of the electric field
is given by
\begin{equation}
(E^{a}_{n,v})^{\dag}=-E^{a}_{n,v}+\Lambda \Xi^{a}_{n,v},
\end{equation}
\noindent where, $\Xi^{a}_{n,v}$, is a linear combination, coming
from different permutations of the three-volume element
\begin{equation}
 \Xi^{a}_{n,v}=(-1)^{a}\left[e^{1}_{n,v}e^{2}_{n,v}\right]e^{0}_{n,v}+(-1)^{a}\left[e^{1}_{n,v},e^{0}_{n,v}
 e^{2}_{n,v}\right]+(-1)^{a}\left[e^{0}_{n,v}e^{1}_{n,v},e^{2}_{n,v} \right].
 \end{equation}
where $a\in \{ 1,2 \}$. With all this at hand, the electric field
along the opposite direction, evaluated at point (n,v) reads
\begin{equation}   \label{eq: E bar a before eq of motion}
E^{\bar{a}}_{n,v}=-(h^{a}_{n,v-e_a})^{\dag}E^{a}_{n,v-e_a}(h^{a}_{n,v-e_a})+\Lambda\
(h^{a}_{n,v-e_a})^{\dag}\Xi^{a}_{n,v-e_a}(h^{a}_{n,v-e_a}).
\end{equation}

\noindent Using the constraints, (\ref{eq: general solution
traces}), and the definition of the electric field, one can show
that
\begin{eqnarray}
  E^{1}_{n+1,v}    &=& +V_{n,v}^{\dag}h^{01}_{n,v}e^{2}_{n,v}V_{n,v} \\
  E^{2}_{n+1,v}    &=& -V_{n,v}^{\dag}h^{02}_{n,v}e^{1}_{n,v}V_{n,v}.
\end{eqnarray}

\noindent Finally, the above relations together with (\ref{eq: E
bar a before eq of motion}), allow us to prove that the equation
(\ref{eq: of motion for Gauss's law}), is equivalent to
\begin{equation}
E^{1}_{n+1,v}+E^{2}_{n+1,v}+E^{\bar{1}}_{n+1,v}+E^{\bar{2}}_{n+1,v}=0
\end{equation}

\noindent This is, of course, the  Gauss's law defined on the
lattice context. When the cosmological constant is present, we
observe that the lattice version of $(2+1)$-gravity, preserves a
remnant symmetry from the continuum theory, this local discrete
symmetry is generated by the internal group thorough the Gauss's
law in its discrete version.\\
From equation (\ref{eq: general solution traces}), we noted that
the holonomy along the spacial plaquettes is proportional to the
cosmological constant, which depending on its value we will have
(Anti)de Sitter solution or as it was considered at \cite{Sigma},
flat solutions when the cosmological constant vanishes. But even
more, the discrete theory admits more solutions than the continuum
one, this is  because there is one more  term $\sigma^{bc}$ which
depends on the plaquettes, in the sense that a possible solution
would be a connection that makes the term $\sigma^{bc}$ be +1 on
certain plaquettes and -1 on others. \\
The next step (it will be presented in a future work), will be to
solve the discrete constraints as operators on a appropiate
Hilbert space, and then construct unitary projectors of the
discrete theory onto the physical space of the continuum theory.
An useful idea for constructing the physical Hilbert space is to
average states in an auxiliary Hilbert space over a suitable
action of the gauge group \cite{raq1}, \cite{raq2}, \cite{raq3}.
For a noncompact gauge group the averaging need not converge, but
when the averaging is formulated within refined algebraic
quantisation, convergence on a linear subspace will suffice.
Within this formalism, and the nature of the discretization, it is
possible to show that the group averaging provides considerable
control over quantisation. On the other hand the discretization
technique can be viewed as a new paradigm for dealing with cases
where the continuum theory does not exist \cite{Gambini1},
\cite{Sigma}.

\section{Conclusions}

In this paper we have presented a continuous and discrete
canonical analysis of the $(2+1)$-dimensional gravity plus
cosmological constant. At the lagrangian level we obtained that
the gauge invariance generated by the space-time diffeomorphisms
is given by the Kalb-Ramond transformations. Then, in order to
make a clear-cut comparison with the discrete case, we carried-out
an extended canonical analysis of the continuous theory, where all
fields were considered as dynamical. Within this context, by
examining the constraints in the complete phase space, we obtained
in an explicit form, the generators of the gauge transformations
for all fields present in the action, even if they behave as
Lagrange multipliers, in accordance with the on-shell Kalb-Ramond
transformations. Finally, we present a discrete canonical analysis
based on the so called variational integrators method. Even
though, the discretization breaks the initial gauge freedom, the
theory preserves certain gauge invariance generated by the Gauss's
law, but in this case defined on the lattice context. In this
manner, a quantisation scheme such as the refined algebraic
quantisation will be of wide interest in the treatment of discrete
gauge symmetries, whose implications will be the aim future
investigations.


\noindent \textbf{Acknowledgements}\\[1ex]
The authors acknowledges support from a CONACyT scholarship
(M\'exico), J.E.R.Q. is supported by PROMEP postdoctoral grant.


\end{document}